# The Clouds of Physics and Einstein's Last Query: Can Quantum Mechanics be Derived from General Relativity?

F. Winterberg

*University of Nevada, Reno NV 89557-0058*

**Abstract**

Towards the end of the 19[th] century, Kelvin pronounced as the "clouds of physics", 1) the failure of the Michelson-Morely experiment to detect an ether wind, and 2) the violation of the classical mechanical equipartition theorem in statistical thermodynamics. And he believed that the removal of these clouds would bring physics to an end. But as we know, the removal of these clouds led to the two great breakthroughs of modern physics: 1) The theory of relativity, and 2) to quantum mechanics. Towards the end of the 20[th] century more clouds of physics became apparent. They are 1) the riddle of quantum gravity, 2) the superluminal quantum correlations, 3) the small cosmological constant. Furthermore, there is the riddle of dark energy making up 70% of the physical universe, the non-baryonic cold dark matter making up 26% and the very small initial entropy of the universe. An attempt is made to explain the importance of these clouds for the future of physics. Conjectures for a possible solution are presented. They have to do with Einstein's last query: "Can quantum mechanics be derived general relativity", and with the question is there an ether?

## 1. Introduction

One great scientific event of the 19[th] century was Darwin's theory of evolution. The theory made a lot of sense provided a sufficient amount of time was given. But as Kelvin, at his time a towering authority in physics, had pointed out (by the theory he and Helmholtz had developed), the sun could not be more than 100 million years old, explaining its radiative energy output to result from the conversion of gravitational potential energy into heat through its shrinking to a smaller diameter. The conclusion by Kelvin was a big blow to Darwin's theory, because it was difficult to believe that a time of 100 million years was long enough to explain the evolution from bacteria to homo sapiens. Today, of course, we know that by the release of energy from nuclear fusion, the sun has a lifetime of about 10 billion years, sufficiently long for Darwin's theory of evolution. This is a good example of how careful one should be in drawing such far reaching conclusions as were drawn by Kelvin, and as they are drawn today with our present (most likely still) limited knowledge of the physical universe. We are though fairly certain, that ordinary matter, of which we and all the stars are composed, makes up only 4% of the material content of the entire universe, with 26% made up from completely unknown heavy particles (the non-baryonic cold dark matter), and 70% of an unknown energy ("quintessence") not even made up of particles.

When Kelvin tried to bring the house of physics in order, not only with regard to the time scale he gave Darwin for his theory of evolution, it was his opinion that apart from a few "clouds", physics is an almost completed science. These clouds were: 1) The failure of the Michelson-Morely experiment to detect an ether wind, and 2) the failure of classical statistical thermodynamics to explain the specific heat of solids at low temperatures. As we know today the removal of these clouds led to the two great breakthroughs of $20^{th}$ century physics: The theory of relativity and quantum mechanics. Adding the discovery of nuclear energy, not anticipated by Kelvin, there were actually all together three clouds.

At the turn from the $20^{th}$ into the $21^{st}$ century there again appear new clouds over the horizon, not less mystifying than were Kelvin's clouds. Drawing from Kelvin's experience, it is very unlikely that these clouds can be removed by the extrapolation of present theories, like the extrapolation of general relativity from the four space-time dimensions of every physics laboratory, to more than four space-time dimensions, and from zero-dimensional point particles to higher dimensional brane particles. What is mathematically obvious, as are these extrapolations, is physically most likely wrong. We must rather look for radical new ideas, such as suggested by Einstein's last query: "Can quantum mechanics be derived from general relativity?" In paraphrasing Bohr (with regard to Heisenberg's failed unified field theory), it is not that at first glance this conjecture seems to be crazy, but if it is crazy enough to be true.

**2.The New Clouds**

At the turn of the 21$^{st}$ century, the most puzzling cloud of physics was and still is the unsolved problem of quantum gravity: How to quantize Einstein's gravitational field equations. Of the four fundamental forces, the electromagnetic, the weak, the strong, and the gravitational force, only the first three of these can be quantized in a mathematically consistent way, but this turns out to be very difficult for gravity. The problem seems to be solvable by making such outlandish conjectures that there are more than the three dimensions of space, and that the point-like structure of elementary particles, (required by the postulates of the special theory of relativity), must be replaced by strings or higher dimensional surfaces (membranes). In making these conjectures, the followers of this line of thought seem to have forgotten that points, lines, and surfaces are abstract elements of Euclidean geometry, having no place in physics, where everything must be measurable. The argument in favor of strings is that the infinites in relativistic quantum mechanical calculations can there be avoided. But because of Heisenberg's uncertainty relation this is an illusion, because to measure the vanishing diameter of a string, or the thickness of a membrane requires an infinite amount of energy. For point-like particles, quantization works for renormalizable theories, like quantum electrodynamics, where the subtraction of two infinities is set equal to the value of an observed quantity. This renormalization "trick" does not work for quantum gravity.

While for quantum gravity the problem is "mathematical", it is for the superluminal quantum correlation (EPR experiment) "physical". About this problem, countless papers and books have been written, making all sorts of philosophical conjectures, with the most outlandish one the "many worlds interpretation". The reason for the failure to find a reasonable explanation for these superluminal connections can be expressed by the formula: Einstein cannot be wrong and there are no superluminal connections. But the experiments, taken at face value, clearly demonstrate the existence of such superluminal connections. The easiest way to explain these superluminal connections is that there is an ether, which in addition to the transmission of electromagnetic and gravitational waves, permits the transmission of superluminal signals. But if there is an ether, it may perhaps also lead to a solution of the problem of quantum gravity, in a much more plausible way than assuming the existence of higher dimensions. As it was shown prior to Einstein by Lorentz and Poincare, the special theory of relativity can be completely and consistently derived assuming the existence of an ether. And with the superluminal quantum correlations, exposing themselves in such an unambiguous way as demonstrated in Aspect's EPR experiment, there should be a way to detect the ether by an experiment.

Two other mysteries of 21$^{st}$ century physics, the non-baryonic cold dark matter, and the dark energy making up 70% of the physical universe, also point in the direction for the existence of an ether. Finally, the small cosmological constant and the very small initial entropy may be connected to each other, suggesting the hidden existence of negative, masses. Assuming that there is an ether, these negative masses must be an essential part of this ether.

### 3. Einstein's Last Query

It is in seeking an answer to Einstein's last query: "Can quantum mechanics be derived from general relativity?" where we may find a key for the solution of all these mysteries. At a first glance, Einstein's query seems to be absurd: How can quantum mechanics be derived from a classical field theory? But we must keep in mind that Einstein's field equations are not only nonlinear, but self-sourced, unlike classical fluid dynamics which is nonlinear but not self-sourced. Self-sourced means, that the gravitational field, having energy and hence mass, is a source of its own field.

This then raises the question: Does one really need in addition to Einstein's vacuum field equation

$$R_{ik} = 0 \qquad (1)$$

the geodesic equation of motion for a mass point singularity

$$\delta \int ds = 0 \quad ? \qquad (2)$$

inasmuch as the equation of motion for a singularity should be a particular solution of the vacuum field equation.

In 1927 Einstein and Grommer [1] tried to solve this problem, and they were able to find a solution in the limit of weak fields. It turned out that in this limit they obtained the classical Newtonian equation of motion of a point particle, and the answer to Einstein's query seemed to be negative. However, this conclusion was only valid under the restriction that the singularity is a mass monopole. This leaves open the question, if something like quantum mechanics can be obtained admitting mass multipoles, containing negative masses, in particular mass dipoles, including the superposition of a mass monopole with a mass dipole (pole-dipole particle). One of the characteristics of quantum mechanics is the "Zitterbewegung" (quivering motion) of a particle. It was shown by Mathisson [2] and by von Weyssenhoff [3] that the motion of a more general mass multipole, strikingly resembles the quivering motion of an elementary particle, first shown by Schrödinger in his famous papers on the "Zitterbewegung" [4] to be an intrinsic property of a Dirac particle. Neither the quantum potential, derived from Schrodinger's equation, nor the motion of a mass multipole obeys the equivalence principle. This, of course, cannot explain all of quantum mechanics, but it is a hint that the existence of negative masses may play an important part in the fundamental laws of nature.

A second important characteristic of quantum mechanics is the superluminal wave function collapse, and a third characteristic is the wave function entanglement.

Entanglement actually already occurs in classical mechanics in the interaction of coupled harmonic oscillators. It there means that an assembly of such oscillators can be decomposed in a set of uncoupled oscillators (normal modes) by a principal axis transformation. Therefore, if as Schrödinger believed, the behavior of a single particle can be described by a wave equation, it can for two particles be described by a set of different quasiparticles (normal modes), each of them described by a wave function not entangled with the wave functions of the other quasiparticles. This description, however, requires to give up the particle aspect entirely, viewing particles not more than the normal modes of an ether. Giving up the multi-particle aspect and replacing it by a set of normal modes in an ether, means that the particles cannot be localized. As waves they are spread out over space. The particle property then only emerges if a measurement is made, by which the wave function collapses into a small volume of space, giving the appearance of a particle, making a spot on a photographic plate for example. This raises the question: What causes the collapse of the wave function? I offer the following conjecture as an answer: It is a kind of gravitational collapse, on the time scale (G Newton's constant)

$$t = (G\rho)^{-1/2} \tag{3}$$

where in an ether with positive and negative masses, the fluctuation of the masses can become very large, with t becoming very small. Under this assumption the collapse velocity

$$v = r/t \propto \rho^{1/2} \tag{4}$$

where $\rho$ is the density of the fluctuating ether. If the special theory of relativity is considered a dynamic ether theory, with electromagnetic and gravitational waves moving with the velocity of light through an ether, as in the pre-Einstein theory of relativity by Lorentz and Poincare, the velocity of light can be surpassed during a gravitational collapse, where in reaching the event horizon, matter held together by electromagnetic or likewise acting forces, becomes unstable and disintegrates into more fundamental particles subject to Newtonian nonrelativistic mechanics. According to our present view these ultimate particles can only be Planck mass particles. And in assuming that there are positive and negative masses, this must be positive and negative Planck mass particles.

**4. The Planck Ether Hypothesis**

To construct a model of the vacuum composed of positive and negative Planck mass particles, we assume that the vacuum is filled with an equal number of positive and negative Planck mass particles with each Planck length volume occupied in the average by one Planck mass particle. And to be consistent with Planck's hypothesis, these Planck mass particles must interact over a Planck length by the Planck force $c^4/G$. One then only has to choose the sign of the force in such a way that the ether composed of positive and negative Planck mass particle is stable. This condition requires that the force between Planck mass particles of equal sign is repulsive, and between those of opposite sign attractive. While the law for the conservation of energy is conserved during the collision between a positive with negative Planck mass particle of mass $m_p$, this is not the case for the momentum which fluctuates by $\Delta p = m_p c$. Therefore, Heisenberg's uncertainty principle

$$\Delta p \Delta q \cong h \qquad (5)$$

where $\Delta p = m_p c$, and $\Delta q = r_p$ ($r_p$ Planck length), is in the Planck ether established at the most fundamental scale, the Planck scale, solely from the hypothesis for the co-existence of positive with negative Planck masses forming a positive-negative Planck mass plasma, or what one might call a Planck ether.

A fluctuation of the momentum, of course, violates Newton's third axiom: actio = reactio. But that Newton's third law is violated in quantum mechanics is established in the sense that the wave function acts on the particle, but not the particle on the wave function.

## 5. The Quasiparticle Spectrum of the Planck Ether

The proposed Planck ether can be compared with a superfluid, but a superfluid possessing two mass components, with the two components interacting in a way that the momentum fluctuates. As it turns out, this medium exhibits a spectrum of elementary particles, greatly resembling the spectrum of particles of the standard model.

Let us first make a comparison with the ground state spectrum of a superfluid, shown in Fig.1, [5]. It is assumed that the spectrum of a superfluid is universal, with its shape taken from measurements of superfluid helium, with the Debye length replaced by the Planck length. In this spectrum the phonons (dilatons) present a scalar gravitational field. The rotons can be identified with the particles of the non-baryonic cold dark matter. The rotons, with an energy somewhat below the Planck energy $m_p c^2$, are separated from the ground state by an energy gap. The magnitude of this gap is in fairly good agreement with the observation that 70% of the energy of the universe does not consist of particles, with the remaining 26% consisting of non-baryonic cold dark matter, identified as heavy rotons, with a mass somewhat smaller than the Planck mass.

Above the ground state, there are higher excitations of the Planck ether. As in a superfluid they result from quantized vortices. There are two such excitations, one in which the two superfluid components co-rotate, and one where they counter-rotate see Fig.2, [5]. In this model the origin of charge is explained to result from a field of virtual phonons, having their source in the zero-point fluctuations of Planck mass particles bound in the vortex fragments, as shown in Fig.3, [5].

With no energy required for the excitation of a large number of positive and negative mass vortices, one obtains by spontaneous symmetry breaking a lattice of vortex rings as shown in Fig.4, [5], with two kinds of waves, both propagating with the velocity of light through this vortex lattice. As shown in Fig.5, [5], one of these waves simulates electromagnetic, and the other one gravitational waves. This reduces the problem of quantum gravity (but also of quantum electrodynamics) to a non-relativistic many body problem.

It was Schrödinger who showed that Dirac's equation implies the existence of negative masses, which by forming a bound state with positive masses makes up for a mass-pole-dipole configuration, possessing a luminal "Zitterbewegung". Vice verse, one can obtain the Dirac equation assuming the existence of negative masses. In the proposed Planck ether model, the Dirac spinors are quasiparticles made up from the positive and negative mass vortex solutions. And with the two kinds of vortex solutions, co- and counter-rotating, there are two families of Dirac spinors as in the standard model. Because the interaction between the vortices is nonlinear, there are no more than four such quansiparticle solutions as the four families in the standard model [5].

The masses of these Dirac spinor particles are explained by the positive gravitational interaction energy between a very large positive and a very large negative mass of the vortex quasiparticles, making up a Dirac spinor. The masses thusly obtained, are small in comparison to a Planck mass, and in fairly good agreement with the masses of the known elementary particles.

This model reproduces all of the symmetries of the standard model, replacing strings in higher dimensions with vortices in the three dimensions of space, and interpreting

Lorentz invariance as a dynamic symmetry for objects held together by electromagnetic or likewise acting forces.

## 6. The Problem of the Small Cosmological Constant and of the Small Initial Entropy

The assumption for the existence of hidden negative masses also sheds light on the two remaining mysteries: The small cosmological constant and the small initial entropy. The zero point energy of the vacuum should lead to a huge the cosmological constant of the order $\Lambda \approx 10^{66}$ cm$^{-2}$. According to Einstein's gravitational field equations, this would mean that the radius of the universe is equal to R = $\Lambda^{-1/2}$ = $r_p \approx 10^{-33}$ cm. But astronomical observations rather suggest that R $\approx 10^{28}$ cm. The cosmological constant is therefore off from what theory predicts by about 122 orders of magnitude. But if there is an equal number of positive besides negative Planck masses making up the Planck ether, both contributing by their mutual Zitterbewegung mechanics to the zero point energy of the vacuum, the cosmological constant would actually have to be zero. The observed small positive value of the cosmological constant can then possibly be explained by a dynamic symmetry breaking of the other-wise "symmetric" positive-negative mass Planck ether.

The prospect to make plausible the smallness of the cosmological constant in the proposed positive-negative mass two-component ether model gives hope to make plausible the very small initial entropy of the universe [6], without invoking the help of God [5]. There we make the following hypothesis: If an assembly of positive and negative masses, with their total energy equal to zero, is brought together, the temperature and hence entropy, of the mixture must go to zero. This hypothesis is the only one consistent with Nernst's theorem, which says that at zero temperature the entropy should be zero. To satisfy this hypothesis, we assume that negative masses have a negative entropy, because only then is an analytical continuation of the entropy from positive to negative temperatures possible.

For the entropy of a mixture of positive and negative masses to become zero in the course of a gravitational collapse requires an exact correlation in the disorder of the positive with the negative masses. This is certainly true if the negative mass is equal to the negative gravitational field energy of the positive mass, because the Newtonian gravitational field of each particle, all the way down to the smallest dimension, is precisely correlated to the position of the particle. The entropy of the positive mass of matter and the entropy of the negative mass of its gravitational field might therefore be called complementary, like a positive and negative photographic image. Then, if in a gravitational collapse these two images exactly overlap, the entropy should go to zero. For the universe, the time needed to bring it back to a low entropy is the Poincare recurrence time. While under normal conditions this time is huge, it may in a dense mixture of positive and negative masses with a divergent acceleration become quite small.

Applied to the universe, this means that following its gravitational collapse, there can be a new beginning with a low entropy.

**7. How Can We Detect the Ether?**

All the talk about the existence of a hypothetical ether must remain empty unless we can find a way for its detection.

I will focus on two possibilities:

1. The breaking of the quantum correlations by a turbulent ether.
2. The influence of strong gravitational fields on the correlations.

To 1.

In terms of the wave number k, the spectrum of the quantum mechanical zero point vacuum energy is given by

$$f(k) = hck^3 \qquad (6)$$

while the spectrum of turbulence goes as

$$F(k) = const.k^{-5/3} \qquad (7)$$

It follows that for sufficiently small wave numbers (long wave lengths), turbulence will overtake the zero point energy fluctuations, and it is conjectured that in crossing this length, the quantum correlations in EPR type experiments begin to break down [7, 8]. To know more exactly at which wave length, the constant in (7) must be known. Furthermore, if this length is different in different directions (against the stellar system, resp. ether rest frame), an absolute ether rest frame can be established, because unlike (6), (7) is not Lorentz invariant.

To 2.

With the wave function collapse conjectured to be some kind of gravitational collapse, it is conceivable that a gravitational field may influence the collapse, for example in the radioactive decay constant in probes brought near to the sun. Most of these highly interesting experiments can be done in the vacuum of space, using the solar system as a laboratory. Ideally, such experiments require a nuclear propulsion system to move large payloads with high velocities through the solar system. This could be done with a chain of exploding mini-nukes. I could think of no better way for mankind to get rid of its large arsenal of nuclear weapons in this way.

**References**

1. A. Einstein and J. Grommer, Preuss. Aksdemic der Wissenschaften, Sitzungsbericht Jan. 6, 1927.
2. M.Mathisson, Acta Physica Polonica 6, 218 (1937)
3. J. von Weyssenhoff, Max-Planck-Festschrift, Deutscher Verlag der Wissenschaften, Berlin 1958.
4. E. Schrödinger, Berl. Ber. 1930, 416; 1931, 418.
5. F. Winterberg, The Planck Aether Hypothesis, Gauss Press, 2002; P. O. Box 18265, Reno, Nevada, 89511, USA; Z. f. Naturforsch. 58a, 231 (2003).
6. R. Penrose, The Empeor's New Mind, Oxford, University Press, 1989.
7. F. Winterberg, Z. Naturforsch. 53a, 659 (1998).
8. E. Gkioulekas, Int J. Theor Phys (2008) 47: 1195-1205.

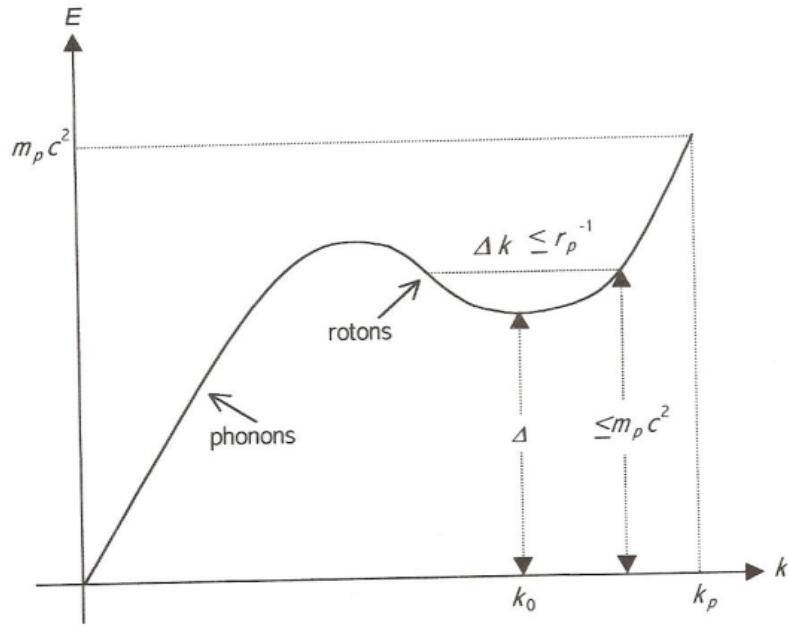

Fig.1. The phonon-roton energy spectrum of the hypothetical Planck ether

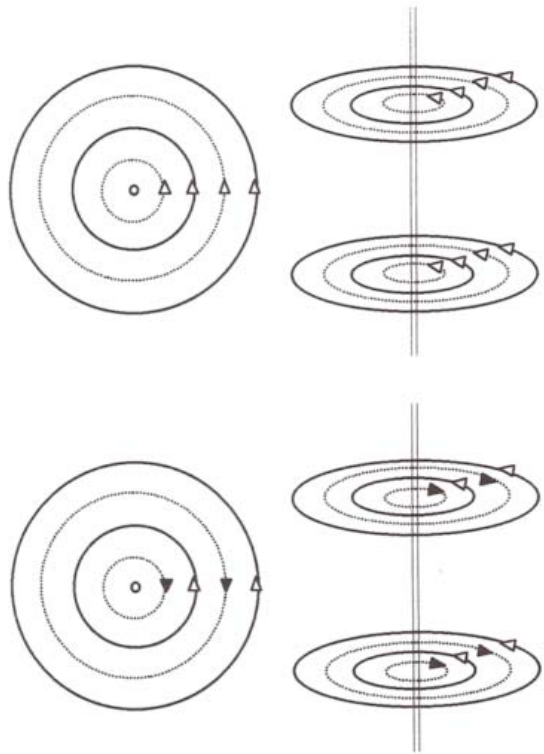

Fig.2. Flow pattern of co- and counter-rotating vortex solutions.

Solid and dotted lines represent the flow of the positive and of the negative Planck mass fluid.

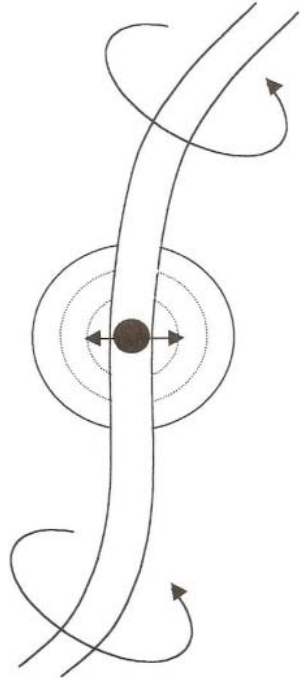

Fig. 3. Origin of charge by zero point oscillations of Planck masses bound in vortex filaments generating a virtual field of phonons, resulting in a long range Newtonian gravitational force.

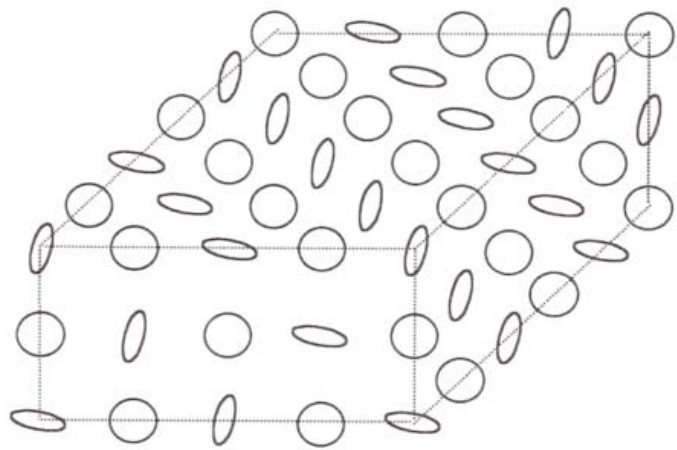

Fig.4. The likely structure of the ring vortex lattice.

A clipping of a rectangular parallelepiped. The orientation of the rings is periodical.

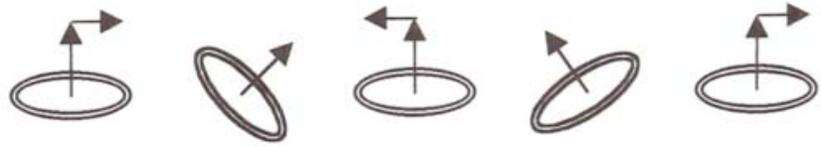

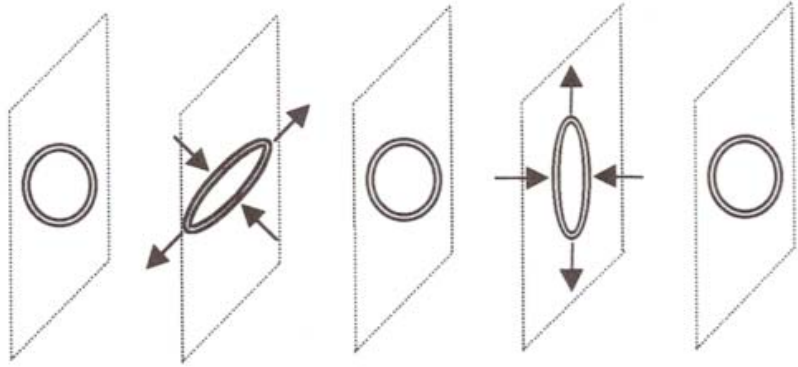

Fig.5. Deformation of the vortex lattice for an electromagnetic and a gravitational wave.